\documentclass[pdflatex,twocolumn,epjc3]{svjour3}          

\RequirePackage[T1]{fontenc}

\smartqed  

\RequirePackage{graphicx}
\RequirePackage{mathptmx}      
\RequirePackage{flushend}
\RequirePackage[numbers,sort&compress]{natbib}
\RequirePackage[colorlinks,citecolor=blue,urlcolor=blue,linkcolor=blue]{hyperref}

\usepackage{epstopdf}

\journalname{Eur. Phys. J. C}

\begin{document}

\title{First design of a crystal-based extraction of 6 GeV electrons for the DESY II Booster Synchrotron}

\author{A. Sytov \thanksref{e1,addr1} \and G. Kube \thanksref{addr2} \and L. Bandiera \thanksref{addr1} \and P. Cirrone \thanksref{addr3} \and H. Ehrlichmann \thanksref{addr2} \and V. Guidi \thanksref{addr1,addr4} \and V. Haurylavets \thanksref{addr5} \and M. Romagnoni \thanksref{addr1} \and M. Soldani \thanksref{addr1,addr4} \and M. Stanitzki \thanksref{addr2} \and M. Tamisari \thanksref{addr6} \and V. Tikhomirov \thanksref{addr5} \and K. Wittenburg \thanksref{addr2} \and A. Mazzolari \thanksref{addr1}}

\thankstext{e1}{e-mail: sytov@fe.infn.it}

\institute{INFN Ferrara Division, Via Saragat 1, 44124 Ferrara, Italy \label{addr1}
          \and
          Deutsches Elektronen-Synchrotron DESY, Notkestr. 85, 22607 Hamburg, Germany \label{addr2}
          \and
          INFN Laboratori Nazionali del Sud, Via Santa Sofia 62, 95123 Catania, Italy  \label{addr3}
          \and
          Dipartimento di Fisica e Scienze della Terra, Universit\`{a} degli Studi di Ferrara, Via Saragat 1, 44124 Ferrara, Italy \label{addr4}
          \and
          Institute for Nuclear Problems, Belarusian State University, Bobruiskaya 11, Minsk 220030, Belarus \label{addr5}
          \and              
          Dipartimento di Neuroscienze e Riabilitazione, Universit\`{a} degli Studi di Ferrara, Via Luigi Borsari 46, 44121 Ferrara, Italy  \label{addr6}
}

\date{Received: date / Accepted: date}

\maketitle

\begin{abstract}
A proof-of-principle experimental setup for the extraction of 6 GeV electrons from the DESY II Booster Synchrotron using the channeling effect in a bent crystal is elaborated. Various aspects of the experimental setup were investigated in detail, such as the particle beam dynamics during the extraction process, the  manufacturing and characterization of bent crystals, and the detection of the extracted beam. In order to optimize the crystal geometry, the overall process of beam extraction was simulated, taking into account the influence of radiation energy losses. As result it is concluded that the multi-turn electron beam extraction efficiency can reach up to 16 \%.

In principle this crystal-based beam extraction technique can be applied at any electron synchrotron in order to provide multi-GeV electron beams in a parasitic mode. This technique will allow to supply fixed-target experiments by intense high-quality monoenergetic electron beams. Furthermore, electron/positron crystal-based extraction from future lepton colliders may provide an access to unique experimental conditions for ultra-high energy fixed-target experiments including searches for new physics beyond the Standard Model.
\end{abstract}

\section{Introduction}

Test beam and irradiation facilities are key infrastructures for detector development in High-Energy 
Physics (HEP). They are dedicated to the qualification of particle 
detectors, materials, and components prior to their installation in HEP experiments. In Europe there exist three major facilities: the DESY II Test Beam Facility at DESY (Hamburg, Germany) with secondary electron or 
positron beams between 1 and typically 6 GeV \cite{4} which is based on the DESY 
II booster synchrotron \cite{5}, the 
DAFNE Beam-Test Facility (BTF) at the Frascati National Laboratory  of INFN 
(LNF) which provides electron or positron beams with tuneable energy from 30 MeV 
to 800 MeV \cite{1,2,3} and the PS East and the SPS North Area at CERN which 
provide primary and secondary hadron and lepton beams between 1 and 400~GeV.

The DESY II synchrotron \cite{DESYII} is used today as the injector for the 3$^{rd}$ generation 
synchrotron light source PETRA III, ramping up single bunches of about 10$^{10}$ 
electrons from $E_{min}$ = 0.45~GeV to typically $E_{max}$ = 6.3~GeV, with a 
possible maximum of 7~GeV. In between two successive fillings which takes 
typically several minutes in top-up operation, DESY II is used parasitically for test beam operation.

At present these test beams are generated in a double conversion process instead 
of using a direct extraction of the primary beam. As a first step a 
bremsstrahlung beam is generated by a 7 $\mu$m thick carbon fibre target which 
is placed in the circulating electron beam of DESY II. As a second step these 
photons hit a converter  target (metal plate), such generating electron/positron 
pairs which are selected by species and momentum depending on the polarity and the strength of the magnetic field of the subsequent dipole magnet spectrometer. Using such internal target, a low rate of secondary particles can be produced 
at every bunch crossing, fulfilling the test beam user requirement for a low 
multiplicity beam. A disadvantage however is the low integrated rate. With three 
internal targets installed at different locations in the ring, the facility 
offers three independent beamlines for user operation. Since its inception and 
start of operation in 1987, the usage of the DESY II Test Beam Facility has 
continuously increased. Meanwhile the EU has supported both access and 
enhancements to the facility within several European grants as e.g. the Horizon 
2020-AIDA2020 one \cite{6}.

\begin{figure}
	\resizebox{83mm}{!}{\includegraphics{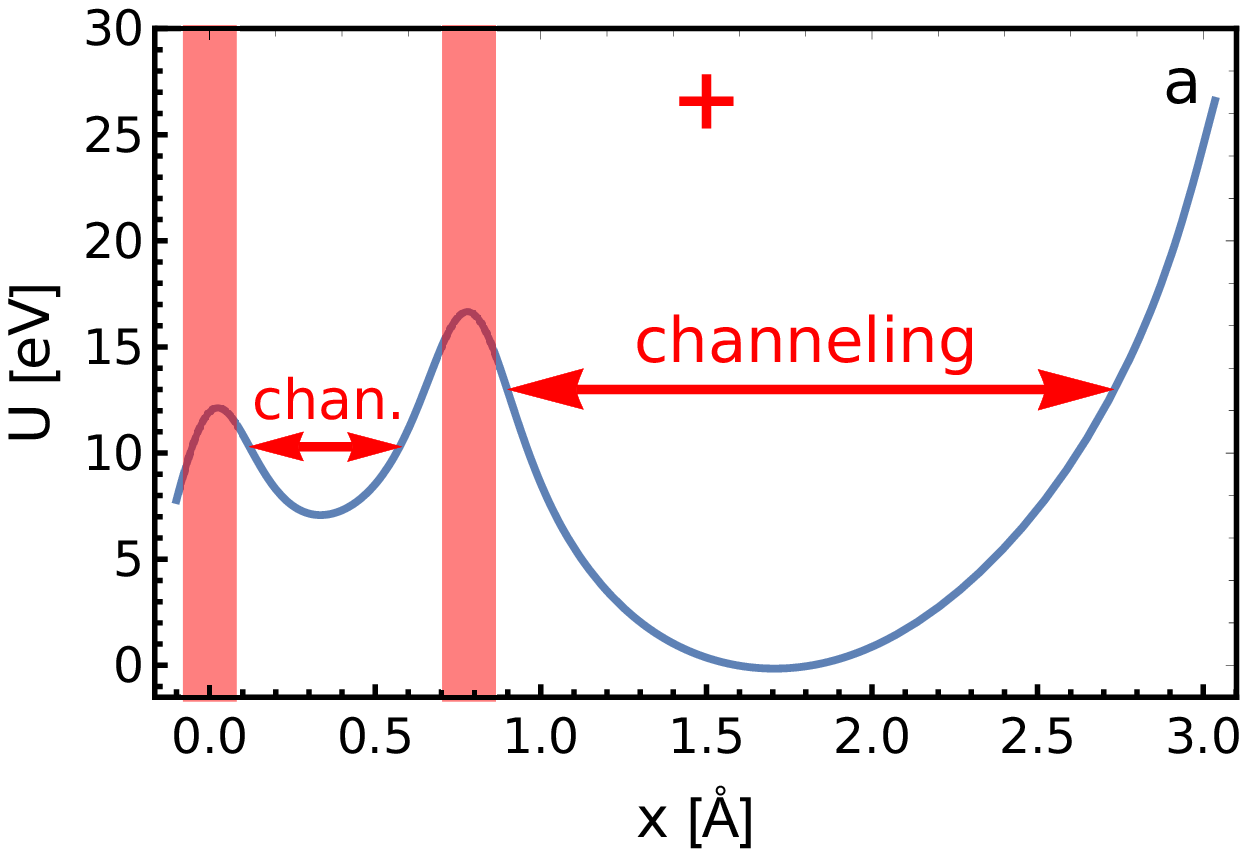}}%
	\\
	\resizebox{83mm}{!}{\includegraphics{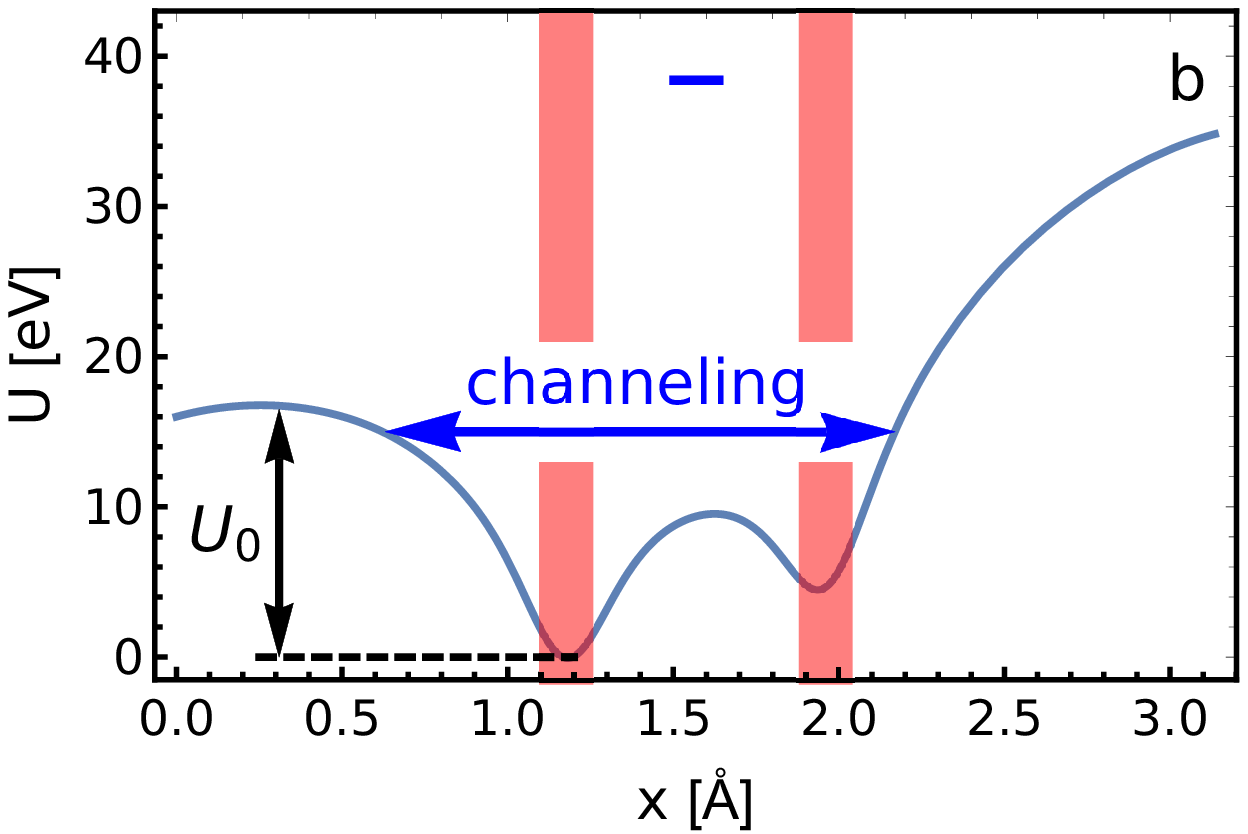}}
	\caption{\label{FigI1} Channeling region (indicated by horizontal arrows) in the interplanar potential formed by (111) silicon bent crystal planes for particles with charge +1 (a) and -1 (b). The vertical lines indicate the zone close to the crystal planes at which Coulomb scattering is more intense. Calculations were done for 6 GeV charged particles and a bending radius of 10 cm. $U_0$ represents the potential well depth
	}
\end{figure}

A significant drawback of the present test beam generation mechanism is the finite lifetime of 
the carbon fibre targets. The experience from wire scanners used in particle 
beam diagnostics shows that thin wire targets may easily be destroyed. The possible reasons are thermal heating of the wire due to energy loss inside the material, higher order mode heating due to vagrant electromagnetic fields inside the vacuum chambers, 
sublimation of the target material, and mechanical stress on the wire fork caused e.g. by a movement of the target for alignment purposes \cite{7}. 
The operational experience with the 7 $\mu$m thick wires of the DESY 
II Test Beam Facility confirms a high risk of target damage which requires a 
frequent wire replacement of 3-4 wires per station and year. Having in mind the upgrade of the synchrotron light 
source PETRA III into an ultralow-emittance source PETRA IV being diffraction 
limited up to X-ray energies of about 10~keV \cite{8} which might imply an 
upgrade of the existing DESY II injector synchrotron, this will increase the 
risk of the target wire damage caused by the substantially lower emittance of 
the new booster ring.

To explore an alternative technique to provide test beams and driven by the 
interest of the facility users,  primary beam extraction schemes are under 
discussion. A method used at hadron machines is the slow resonant extraction 
scheme where typically the third-order resonance is intentionally excited by 
controlling the tune distance and sextupole strength in order to gradually peel off particles 
from outer to inner regions inside the beam emittance \cite{9,10}. This method 
is not solely restricted to hadron machines, the scheme has been applied at 
electron accelerators as e.g. pointed out in Ref.~\cite{11}.

Another scheme which will be discussed in the following is crystal-assisted beam 
steering. Since the 1950s, it has been known that the crystalline lattice 
structure can strongly influence the electromagnetic processes. Such coherent orientational effects in a crystal can be 
exploited for various applications in accelerator physics as well as for the 
development of novel X- and gamma-ray crystal-based sources. In the remainder of 
this article, a design proposal for a proof-of-principle experiment is worked 
out in order to study a crystal-based beam extraction at the booster 
synchrotron DESY II. In the future, this scheme could be applied in order to 
realize a primary beam extraction for the DESY II Test Beam Facility.

\begin{figure}
\resizebox{83mm}{!}{\includegraphics{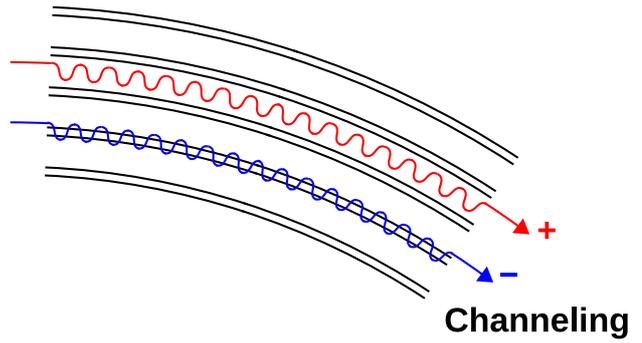}}
\caption{\label{FigI2} Schematic illustration of channeling in a bent crystal formed by non-equidistant crystal planes (e.g. (111)). The signs indicate schematic trajectories either for positively or negatively charged particles
}
\end{figure}

\section{Crystal-assisted beam steering}
The main idea of crystal-assisted beam steering, firstly proposed by Tsyganov in 
1976~\cite{Tsyganov}, relies on planar channeling \cite{Lindhard}. Planar 
channeling is a coherent effect of penetration of charged particles in a crystal 
almost parallel to its planes, when a charged particle is held in a potential 
well (as shown in Fig. \ref{FigI1}) formed by the electric field of two 
neighboring atomic planes. If a crystal is bent (as illustrated in Fig. 
\ref{FigI2}), the charged beam will be steered since its trajectory is confined 
under channeling conditions along the bent crystal planes.

\begin{figure}
\resizebox{83mm}{!}{\includegraphics{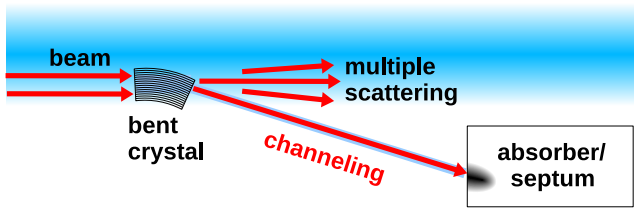}}
\caption{\label{FigI3} Schematic illustration of a crystal-based collimation/extraction. The bent crystal deflects the beam onto either an absorber or a septum magnet. A fraction of the beam is not captured under channeling conditions and will be scattered by the bent crystal}
\end{figure}

Beam steering based on the coherent interaction of charged particle beams with 
bent crystals has found several applications in accelerator physics. In particular, 
crystal-based beam collimation and extraction were successfully investigated at 
several proton synchrotrons, such as the U70, RHIC, Tevatron, SPS and the LHC 
\cite{U70,RHIC,Tevatron,Tevatron2,SPSUggerhoj,UA9,UA92,LHC,LHCion}. A new proposal of slow extraction of positrons from DAFNE has been recently published \cite{DAFNEextraction}. The main conception of both crystal 
collimation and extraction consists in interception of beam halo by a bent 
crystal and consequent deflection under the channeling conditions (as shown in 
Fig. \ref{FigI3}). The difference between collimation and extraction is 
merely only the device onto which the beam halo is deflected, namely 
absorber and septum magnet, respectively.

The main advantage of such a technique is extraction of the beam in a parasitic 
mode, i.e. not disturbing the main fraction of the beam, and, therefore, not 
affecting the main use of the accelerator. Manufacturing and installation of a 
bent crystal is much cheaper than of any kind of magnetic or electrostatic 
deflector. Moreover, the extraction efficiency approaches a 100 \% for positively charged particles. All of this 
makes a bent crystal an ideal solution for beam extraction for fixed target 
experiments at modern accelerators.

Though the crystal-based extraction technique is well developed and has been studied for 
proton beams, it has never been applied for electrons. The main limitation is the significantly lower channeling efficiency for electrons since unlike protons the 
negatively charged particles move under channeling conditions traversing atomic 
planes where Coulomb scattering is considerably stronger (compare Fig. \ref{FigI1}a and b representing the case of positively and negatively charged particles, respectively). Moreover, unlike 
proton synchrotrons usually operating at energies from tens of GeV up to few 
TeV, electron machines usually reach energies of a few GeV, at which the 
contribution of multiple scattering into the channeling process is much more 
important. In addition, the radiation losses for electrons may be an important 
factor, since it may significantly change a particle trajectory in an 
accelerator leading to beam losses.

However, recent experiments at both the Mainz Microtron MAMI (855 MeV electrons) 
\cite{PRL2014,EPJC2017,JINST2018} and the SLAC FACET facility (multi-GeV 
electrons) \cite{PRLSLACold,PRABSLAC,PRLSLAC,NIMBSLAC,ModPhysSLAC} with the new 
generation of bent crystals \cite{JINST2018,Crystals,Crystals2} have demonstrated channeling efficiencies in the 
range from 10 to 40 \%, i.e. of the same order of magnitude as for protons. 
Apart from the ultra-low crystal thickness (tens of microns) and very high 
bending angle (up to few mrad) the crystals possess a very high quality 
crystalline lattice and uniform bending as well.

These crystals were developed not only as beam steering devices, but also as 
innovative X- and $\gamma$-ray radiation sources 
\cite{PRL2015,EPJC2021,PRA2012,NIMB2013,PRL2013,ICHEP2016,PRL2018,BandieraNIMA,
NIMBSLAC,ModPhysSLAC,CU1,CU2,CU3,CU4}, compact electromagnetic calorimeters \cite{PRL2018,BandieraNIMA,KLEVER} and positron sources \cite{Positron1,Positron2,Positron3,Positron4} based on the effect of channeling radiation (CR) \cite{Kumakhov,Baryshevsky}. This type of radiation allows to 
considerably increase the radiation intensity in comparison with bremsstrahlung. 
However, the probability of photon emission in such short crystals should not 
exceed 10-20\%, and, therefore, may not significantly affect the extraction 
efficiency, though must be definitely taken into account. Therefore, such kind of bent crystal should be optimal for the first crystal-based extraction scheme 
for electrons.

The DESY II booster ring is a well suited synchrotron for the first experimental test of the new technique by the following reasons:

\begin{enumerate}

\item The electron beam energy between 450 MeV and 6.3 GeV lies in the energy range already tested in the experiments on channeling 
mentioned above and is typical for electron synchrotrons existing in the world. Such beams are of interest for testing of nuclear and particle physics detectors and generic detector R\&D \cite{4}.

\item The upgrade of the Test Beam Facility consisting in the extraction of a primary, and consequently, low-emittance and very intense electron beam into beam test area is an excellent motivation for these studies.

\item The extraction line including septum magnets already exists. Therefore, in general, only a bent crystal has to be installed for the proof-of-principle experiment.

\end{enumerate}

\section{The DESY II accelerator and beam extraction scheme}

The electron synchrotron DESY II operates mainly as the injector for the 
3$^{rd}$ generation synchrotron light source PETRA III. It accelerates and 
decelerates in a sinusoidal mode with a frequency of 12.5 Hz. The revolution 
frequency is 1 MHz, the RF frequency 500 MHz, and the bunch length is around 30 ps. 
The relevant DESY II beam parameters for the beam halo extraction are summarized in Table \ref{tab:table1}. 

\begin{table}
\centering
\caption{\label{tab:table1} DESY II accelerator parameters \cite{DESYII}}
\label{parset}
\begin{tabular*}{\columnwidth}{@{\extracolsep{\fill}}ll@{}}
\hline
Parameter &Value\\
\hline
Ring circumference $S_0$& 292.8 m\\
Injection energy $E_{min}$& 0.45 GeV\\
Nominal extraction energy $E_0$& 6 GeV\\
Number of $e^{-}$ in the beam $N_0$&  $\sim 10^{10}$\\
Horizontal emittance $\varepsilon_x$ (at 6 GeV)& 339 nm$\;$rad\\
Vertical emittance $\varepsilon_y$ (at 6 GeV)& 35 nm$\;$rad\\
Horizontal tune $Q_x$& 6.7\\
Vertical tune $Q_x$& 5.7\\
Energy spread $\sigma_{\delta E/E_0}$& $0.977\times10^{-3}$\\
Total RF voltage $V_s$& 13.5 MV\\
Harmonic number $h$& 488\\
Gamma transition $\gamma_{tr}$& 6.428\\
\hline
\end{tabular*}
\end{table}

The general lattice layout of DESY II has an eight-fold symmetry, each of the 
eight super periods consists of three FODO cells, and each FODO cell provides a 
2220 mm long drift space. From these twenty-four straight sections, eight are 
equipped with RF-cavities, and the remaining sixteen ones with kicker- and 
septa magnets for injection and extractions and additional installations as e.g. 
the vacuum chambers housing the wire targets for the present test beam 
generation. As mentioned before, for the proof-of-principle experiment an 
existing extraction region can be utilized which was formerly in use for the 
DORIS synchrotron which was in operation from 1974 until 2013. The schematic layout of 
this extraction region is shown in Fig.~\ref{fig4} and will be 
described in the next paragraph.

\subsection{Beam extraction setup}

\begin{figure}[!h]		
	\centering
	\resizebox{83mm}{!}{\includegraphics{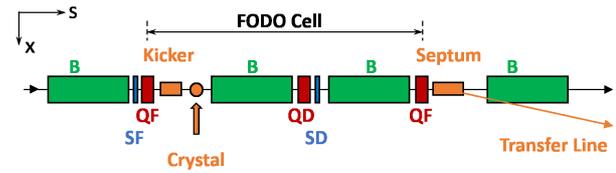}}
	\caption{Scheme of the beam extraction into the former DORIS transfer line. The extraction is embedded in the DESY FODO lattice consisting of dipoles (B) and focusing/defocusing quadrupoles (QF/QD). In addition, the two sextupole families (SF/SD) are indicated. The overall distance from the planned crystal location to the septum position is about 10 m}
	\label{fig4}
\end{figure}

The extraction setup occupies two subsequent straight sections. The first 
section houses the pulsed deflection device that kicks the beam (or in case of 
non-pulsed crystal deflection parts of the beam halo) onto a pulsed septum 
magnet placed downstream. At present it consists of a kicker magnet with a 
maximum deflection angle of 1.14 mrad in positive $x$ direction if supported by 
a beam bump, c.f. Fig.~\ref{fig4}. The second straight section houses the beam 
septum which deflects the separated beam portion into the transfer line, i.e. 
the septum provides space separation between the circulating and extracted beam. The 
extracted beam passes then through the homogeneous field region of the septum, 
the circulating one is in the field-free septum region. In order to provide 
sufficient deflection angle the septum is formed by two individual septum 
magnets each of 600 mm iron length with a 3 mm thick eddy current shielding, 
thus separating the region of homogeneous field from the field-free region for 
the circulating beam. The separated beam portion is then transported along the 
transfer line, its intensity and beam shape can be measured by beam current 
monitors and a scintillator based screen monitor. 

For the planned experiment the deflecting crystal will be installed in the first 
straight section. In principle the crystal could be operated as deflection 
device instead of the kicker magnet. However, in case that the crystal 
deflection is not sufficient, it can be used in order to support the beam kick 
towards the septum magnet.

Before designing a dedicated deflection device, in the subsequent paragraph the required deflection angle is estimated based on the lattice parameters at the  crystal location and the septum magnet. 
\begin{figure*}
	\resizebox{85mm}{!}{\includegraphics{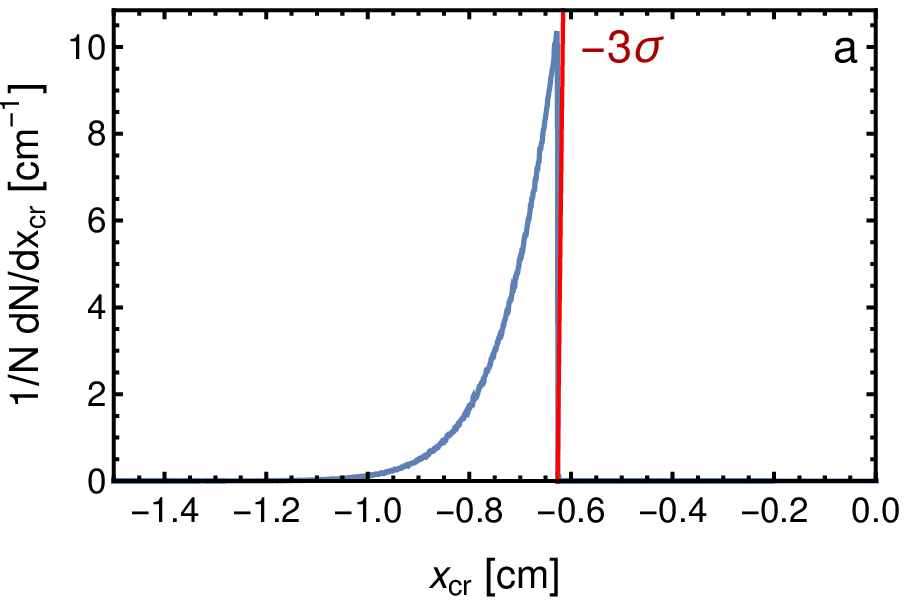}}
	\resizebox{85mm}{!}{\includegraphics{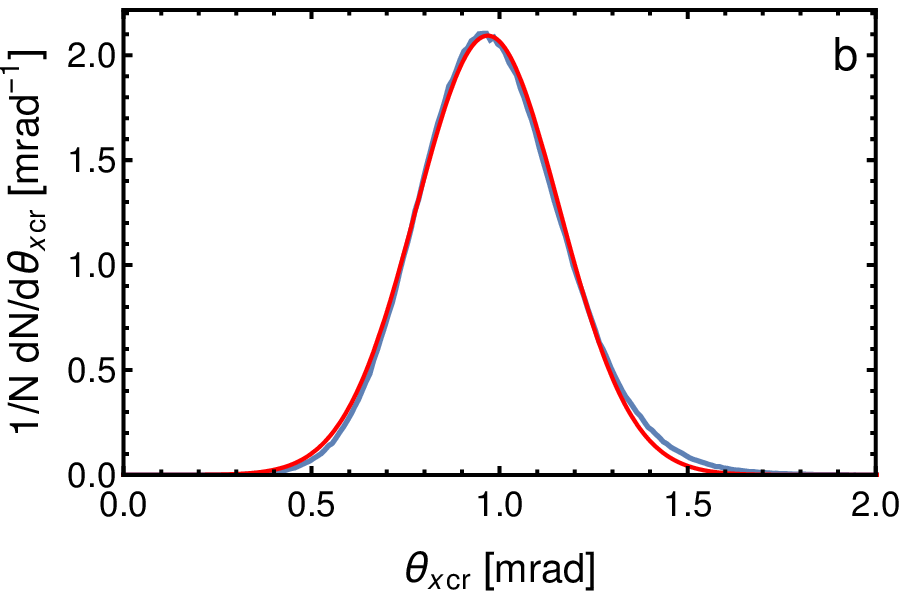}}	
	\caption{\label{Fig1} Transverse (a) spatial coordinate and (b) angular distributions of the beam at the position of the deflection device (bent crystal). The crystal boundary is placed at $x=-3\sigma$.  The red curve in (b) indicates the result of a Gaussian fit}
\end{figure*}

\subsection{Required deflection}

The lattice parameters at the positions of both the deflection device and the septum magnet, which are essential for the estimation of the deflection angle, are summarized in Table \ref{tab:table2}. With these parameters it is possible to estimate the beam position and the angle at the septum based on the beam parameters at the deflection device, i.e. to study the influence of the crystal on the extracted beam.

\begin{table}
\centering
\caption{\label{tab:table2} Optics parameters at the positions of the deflection device and the septum magnet}
\label{parset}
\begin{tabular*}{\columnwidth}{@{\extracolsep{\fill}}lll@{}}
\hline
Parameter&Deflection device&Septum\\
\hline
			Longitudinal position $s$& 148.33 m & 158.89 m\\
			Horizontal $\alpha$-function $\alpha_x$& 1.77 & 2.17\\
			Vertical $\alpha$-function $\alpha_y$& -0.85 & -0.43\\
			
			Horizontal $\beta$-function $\beta_x$& 12.85 m & 17.80 m\\
			Vertical $\beta$-function $\beta_y$& 6.07 m & 4.22 m\\
			
			Horizontal phase advance $\mu_x$& 3.37 & 3.63\\
			Vertical phase advance $\mu_y$& 2.93 & 3.11\\
			
			Horizontal dispersion $D_x$ & 1.60 m & 1.53 m\\
			Vertical dispersion $D_y$& 0 & 0\\
			
			Horizontal dispersion prime $D'_x$ & -0.221 & -0.175\\
			Vertical dispersion prime $D'_y$& 0 & 0\\
\hline
\end{tabular*}
\end{table}

Due to the fact that kicker and septum magnets provide only kicks in the horizontal 
direction, in the subsequent discussion the particle motion will be considered 
only in this plane. Furthermore, it is assumed that the fraction of the beam to 
be extracted is the fraction with horizontal coordinates less than -$3\sigma 
\approx -6.3$ mm of the beam transverse size $\sigma=\sqrt{\varepsilon \beta_x}$. 
This value would be enough to consider the extraction scheme as non-perturbing, 
i.e. stripping away particles with larger horizontal offsets will not affect the 
operation of DESY II as injector for the synchrotron light source PETRA III 
during user operation. While the beam orbit is not stable during the DESY II 
acceleration cycle and moves to the ring inside at higher beam energies, the 
sign in the -$3\sigma$ limit indicates that extraction takes place only at 
higher beam energies for a limited number of turns when the beam has moved far 
enough towards the inner ring side such that there is a spatial overlap between the beam halo 
and the extraction crystal.

The horizontal coordinate $x$ and the angle $x' = \theta_x$ are described by both betatron and synchrotron oscillations as:
\begin{eqnarray}
x & = & x_0 \cos\Psi_x + \delta D_x,
\label{Eq1} \\
\theta_x & = & -\frac{x_0}{\beta_x} (\alpha_x \cos\Psi_x + \sin\Psi_x) + \delta D'_x,
\label{Eq2}
\end{eqnarray}
where $x_0$ and $\Psi_x$ are the amplitude and phase of the betatron oscillations, 
and $\delta=\frac{\Delta p}{p_0}$ is the relative momentum spread with $p_0$ the 
design particle momentum and $\Delta p$ the absolute momentum spread. Assuming 
Gaussian distributions for the amplitudes of both betatron and synchrotron 
oscillations, as well as selecting only the coordinates $x < -3\,\sigma$ 
according to the extraction limit, it is straightforward to generate the distributions of both $x$ and $\theta_x$ at the location of the bent crystal as 
shown in Fig. \ref{Fig1}a-b, respectively. The standard deviation of the Gaussian fit of 
the distribution in Fig. \ref{Fig1}b yields a beam angular divergence at the 
crystal entrance of 0.18 mrad.

The horizontal beam coordinate at the septum can be controlled by the existing 
extraction kicker (for one dedicated revolution) or the beam bump mentioned 
beforehand (for a number of revolutions). In order to choose this coordinate one 
can refer the transverse coordinate distribution at the crystal entrance (see 
Fig. \ref{Fig1}a), which roughly lies between $|$3$\sigma|$ and $|$5$\sigma|$. A 
reasonable choice would be $4 \sigma$. Indeed, taking into account the septum boundary 
thickness being equal to 3 mm (roughly from 4$\sigma$ to 5$\sigma$ at the septum 
magnet position), this will make the entrance into the septum almost unreachable 
for most of the particles without an angular kick by the deflection device. 
Therefore, we define the transverse positions of the deflection device and the 
septum magnet for further consideration to be $-3\sigma$ and $4 \sigma$, 
respectively. However, it should be underlined that both positions may further 
be optimized during the experiment.

In order to define the angular kick, one should be able to transport the 
particles from the location of the deflection device to the location of the 
septum magnet. In other words, one should be able to transform the distributions 
in Fig. \ref{Fig1} into the analogical distributions at the septum magnet 
position using a fixed angular kick. This can be done using the following 
transport equations \cite{HandbookAcc}:
\begin{figure*}
\resizebox{86mm}{!}{\includegraphics{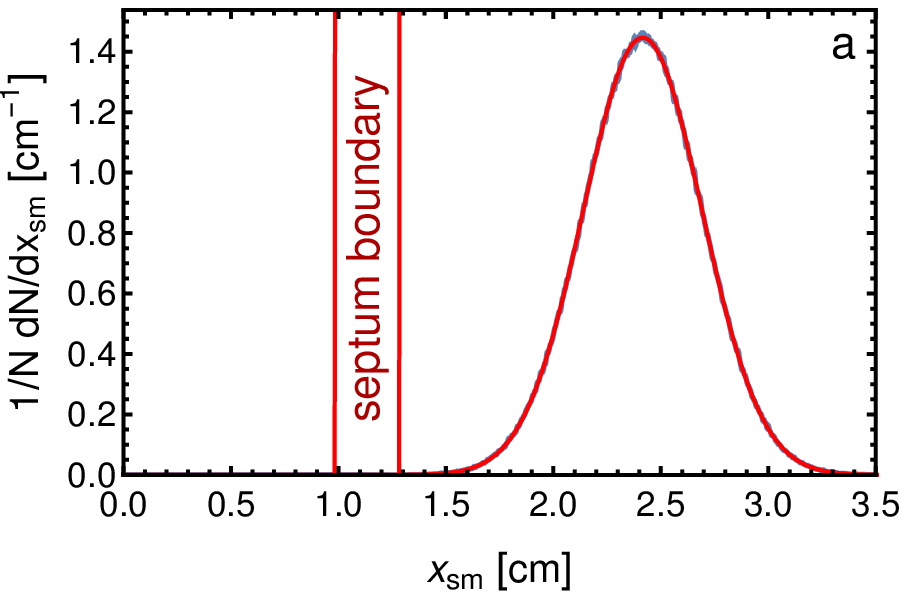}}
\resizebox{86mm}{!}{\includegraphics{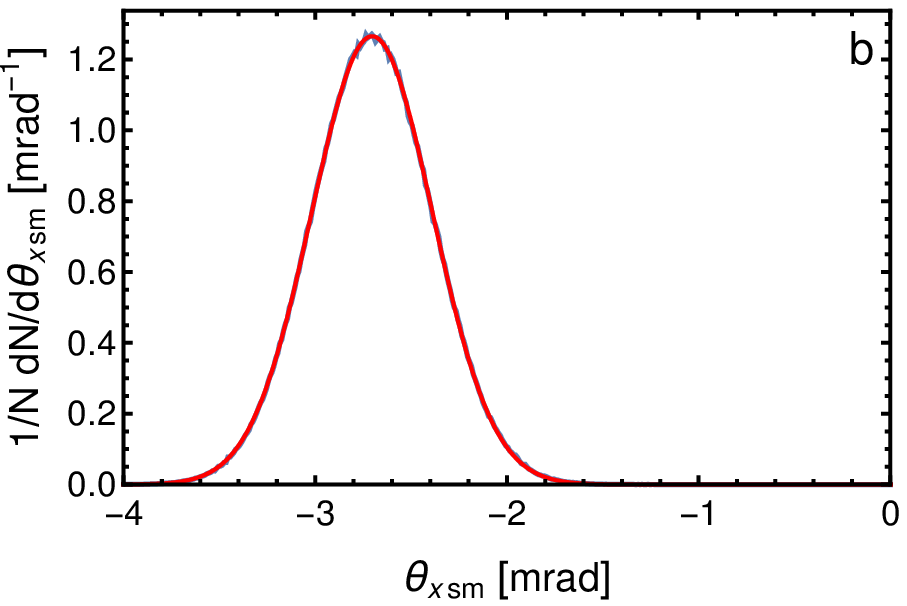}}
\caption{\label{Fig3} Transverse (a) spatial coordinate and (b) angular distributions of the beam at the position of the septum magnet with the condition of a constant angular kick of the deflection device of 1.75 mrad. Vertical lines in (a) indicate the septum magnet boundary. The red curves indicate the results of Gaussian fits}
\end{figure*}

\begin{figure}
\resizebox{83mm}{!}{\includegraphics{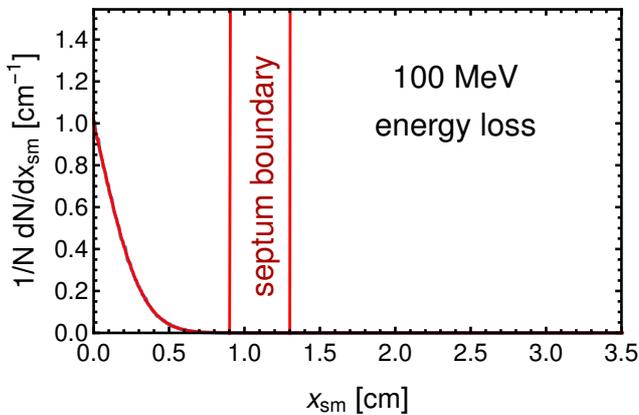}}
\caption{\label{Fig4} Transverse coordinate distribution of the beam at the position of the septum magnet with the condition of a constant angular kick of the deflection device of 1.75 mrad, taking into account the energy loss in the deflection device of 100 MeV. The vertical lines indicate the septum magnet boundary. The red curve indicates the result of a Gaussian fit
}
\end{figure}

\begin{eqnarray}
&x_{sm} = (x_{cr} - \delta D_{x\,cr})\sqrt{\frac{\beta_{x\,sm}}{\beta_{x\,cr}}} \left(\alpha_{x\,cr} \sin \Delta \Psi_x+\cos \Delta \Psi_x \right)+& \nonumber \\
&\left(\theta_{x\,cr} - \delta D'_{x\,cr} + \Delta \theta_x \right)\sqrt{\beta_{x\,cr} \beta_{x\,sm}} \sin \Delta \Psi_x + \delta D_{x\,sm},& \label{Eq4} \\
&\theta_{x\,sm} = -\frac{(1+\alpha_{x\,cr} \alpha_{x\,sm})\sin \Delta \Psi_x + (\alpha_{x\,sm}-\alpha_{x\,cr}) \cos \Delta \Psi_x}{\sqrt{\beta_{x\,cr} \beta_{x\,sm}}}\times&\nonumber \\&(x_{cr} - \delta D_{x\,cr})+ \sqrt{\frac{\beta_{x\,cr}}{\beta_{x\,sm}}}(\cos \Delta \Psi_x-\alpha_{x\,sm} \sin \Delta \Psi_x)\times& \nonumber \\&\left(\theta_{x\,cr} - \delta D'_{x\,cr} + \Delta \theta_x \right) + \delta D'_{x\,sm},&\label{Eq5}
\end{eqnarray}
where the indices $cr$ and $sm$ indicate the position of deflection device 
(crystal) and septum magnet, respectively, and $\Delta \Psi_x$ is the betatron 
phase shift between both locations. $x_{cr}$ and $\theta_{x\,cr}$ are the 
incident spatial coordinate and angle to the deflection device, respectively. 
Considering the deflection device having point-like longitudinal extension, one 
may consider only the change of the angle on the value of the angular kick 
$\Delta \theta_x$.

By using the distributions from Fig. \ref{Fig1} and Eqs. (\ref{Eq4}-\ref{Eq5}) 
one can calculate both the coordinate and angular distribution of the deflected 
beam at the septum magnet entrance. These distributions with the condition of 
the angular kick of 1.75 mrad are shown in Fig. \ref{Fig3}a-b. The mean values 
of these distributions are 2.42 cm and -2.7 mrad respectively. The angular kick 
was chosen to ensure that all the beam being inside $4\sigma$ of the 
coordinate distribution, enters the septum magnet. Some additional gap 
between the septum magnet boundary and the extracted beam (roughly 1$\sigma$ 
thick as shown in Fig. \ref{Fig3}a) could be useful for the adjustment of the 
deflection device and septum magnet positions.

By now we considered an ideal deflection device with a constant angular kick and 
without energy loss as well. However, if one considers a bent crystal, one 
should take into account also radiation energy losses, caused by bremsstrahlung 
or similar effects, since it may cause a considerable trajectory transverse 
shift. In order to take it into account one should substitute the value $\delta$ 
in  Eqs. (\ref{Eq4}-\ref{Eq5}) by the $\delta$ calculated using the energy 
$E-\Delta E$ instead of its initial value $E$, where $\Delta E$ is the radiation 
energy loss value. By putting a typical value of radiation energy loss $\Delta E=100$ MeV covering with some margin both the double height of DESY II RF-bucket and the beam energy increment during a few hundred revolutions in the accelerator, one obtains the following 
coordinate distribution at the septum magnet entrance shown in Fig. \ref{Fig4} 
(to be compared with Fig. \ref{Fig3}a). One can conclude that the radiation loss 
may lead to missing the septum magnet by less energetic particles. Since the 
radiation probability depends on the material thickness, one should make the 
crystal as short as possible. At the same time, a few MeV of energy loss may not 
be so important for the extraction process. Moreover, this influence may be 
reduced by the additional deflection gap mentioned above (Fig. \ref{Fig3}a).

\section{Simulation code}

The CRYSTALRAD simulation code \cite{CRYSTALRAD,CRYSTAL2,RADCHARM} is a Monte 
Carlo code providing fast simulations of both charged particle dynamics and 
radiation spectra in straight, bent and periodically bent crystal of any 
material and crystal lattice type with well verified experimentally models of 
scattering \cite{EPJC2020,EPJC2017,PRLSLAC} and radiation 
\cite{CRYSTALRAD,EPJC2021} at the Mainz Mikrotron MAMI for 855 MeV electrons. The code 
exploits the following approaches.

\subsection{A trajectory in a crystal}
The charged particle trajectory is calculated assuming a relatively small angle with 
respect to the crystal planes or atomic strings in the approximation of an 
averaged atomic potential. It is a quite common approach to simulate channeling 
effects \cite{Lindhard} and has been well validated. A trajectory is calculated 
by a numerical solution of the trajectory equation using a Runge-Kutta 4$^{th}$ order method 
\cite{RungeCutta,Samarskii}. Multiple and single Coulomb scattering on a screened 
(Yukawa) atomic potential as well as single scattering on electrons are randomly 
simulated at each step of the trajectory equation solution. The Coulomb scattering 
model includes only the incoherent part of scattering
\cite{EPJC2020,Tikhscattering}, while the coherent part is simulated by a 
trajectory calculation in the averaged atomic potential.

\subsection{Radiation losses}
The radiation spectrum is calculated using the Baier-Katkov method \cite{Baier}. This 
is a widely used model which uses a classical 
trajectory as an input, but takes into account the quantum recoil of electrons  and positrons in the emission of photons. This model allows to calculate the spectra of channeling radiation as well as coherent bremsstrahlung, i.e. the effects caused by the interaction of charged particles with the ordered crystalline structure. The 
Baier-Katkov method represents itself as a multidimensional numerical integral 
of the simulated particle trajectory and angles of radiation emission direction 
depending also on the value of the radiated energy. In CRYSTALRAD we use the 
Newton-Cotes quadrature rule for the integral of the trajectory as well as the Monte 
Carlo integration of the radiation emission direction. By calculating this integral 
for different values of the emitted energy one obtains the radiation spectrum. This 
spectrum is used for the calculation of the cumulative distribution function, 
which is consequently applied to randomly generate the event of radiation 
emission, and, if it occurs, the radiation energy losses. Hereinafter a 1 MeV 
low-energy cut on radiation production is applied, since a 1 MeV energy loss 
has almost negligle influence on the accelerator beam dynamics in the case of 
DESY II.

\subsection{Accelerator routine}
The accelerator routine \cite{FCCWEEK,VANT} is needed to simulate the particle 
dynamics in an accelerator, and, in particular, the crystal-based collimation or 
extraction scheme. It has been already applied to simulate crystal-based 
collimation of the SPS (120 GeV), LHC (7 TeV) and FCC-hh (50 TeV) 
\cite{FCCWEEK,VANT}. It provides simulations of transverse coordinates and 
angles at certain longitudinal positions in an accelerator, taking into account both betatron and synchrotron oscillations. This routine exploits the transfer equations 
(\ref{Eq4}-\ref{Eq5}) both in the horizontal and in the vertical plane to simulate 
the accelerator optics as well as the numerical solutions of the differential 
equations describing the synchrotron motion \cite{HandbookAcc}. A trajectory 
starts at the crystal entrance surface and finishes at the septum magnet 
entrance or after the set number of turns in the accelerator is exceeded.

\section{Simulations of crystal-based extraction}

\subsection{Channeling effect simulation}

A bent crystal is capable to provide an efficient deflection of the beam halo onto a septum magnet without disturbing the main part of the beam. 
Electron beam deflection becomes possible due to the channeling effect of electrons in a bent crystal. The efficiency of such deflection strongly 
depends on the angular divergence of the beam as well as the crystal geometry, i.e. both the crystal thickness and the bending angle. 

The physical mechanism of this dependence is the following. The channeling effect is possible only at very low incident angles of particles w.r.t. 
bent crystal planes, i.e. less than the critical channeling angle (Lindhard angle) \cite{Lindhard}:
\begin{equation}
\theta_L=\sqrt{\frac{2U_0}{E}},
\label{Eq6}
\end{equation}
where $U_0$ is the depth of the potential well, shown in Fig. \ref{FigI1}b. This angle 
depends on $U_0$ becoming lower for higher values of the crystal curvature. The 
maximal possible value of $\theta_L$ at fixed energy is reached for a straight 
crystal. For the case of a quasi-mosaic Si crystal bent along the (111) planes 
(see Fig. \ref{FigI1}) it is equal to 0.088 mrad, which is less but comparable with the 
angular divergence of the incident beam of 0.18 mrad as calculated above.

The impact of the crystal geometry on the deflection efficiency is determined by two 
competitive effects. On the one hand it is the effect of dechanneling, i.e. particles 
escape from the channeling condition, mainly caused by incoherent Coulomb scattering. 
To minimize this effect, the interaction time between particle and crystal lattice should be kept small, i.e. the crystal should be as short as possible. On 
the other hand, the crystal bending changes the critical channeling angle. Since the 
bending angle is nearly fixed for our task, the crystal 
thickness has to be to increased in order to maximize the channeling acceptance, and consequently, the channeling 
efficiency. This leads to an optimal crystal thickness value.

For a preliminary optimization of the thickness a Si (111) bent 
crystal is chosen since it provides the highest possible efficiency measured 
experimentally for electrons \cite{EPJC2017} in the few-GeV energy range. The 
crystal bending angle has been fixed at the value of $\theta_b=$1.75 mrad as defined above. The incoming beam is the same as in Fig. \ref{Fig1}. The 
simulations of the beam deflection in a bent crystal at different crystal 
thicknesses has been carried out using the CRYSTALRAD simulation code.

\begin{figure}
\resizebox{83mm}{!}{\includegraphics{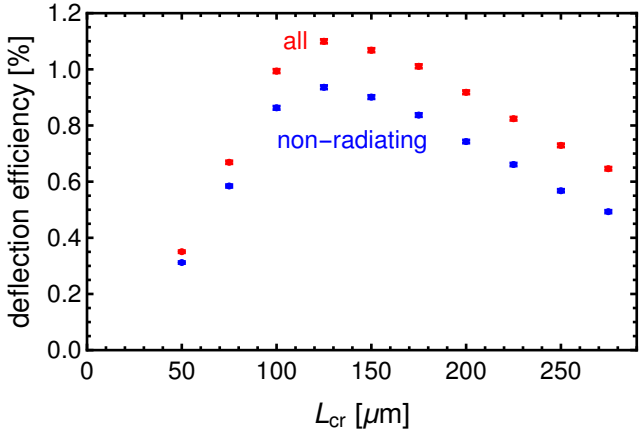}}
\caption{\label{Fig5} Deflection efficiency as function of the crystal thickness for a Si crystal bent along the (111) planes with a deflection angle of 1.75 mrad. The lower blue dots represent the fraction of channeled particles that do not emit radiation, while for the upper red points the radiation process is possible. The incoming beam distribution is the same as in Fig. \ref{Fig1}}
\end{figure}

\begin{figure}
\resizebox{83mm}{!}{\includegraphics{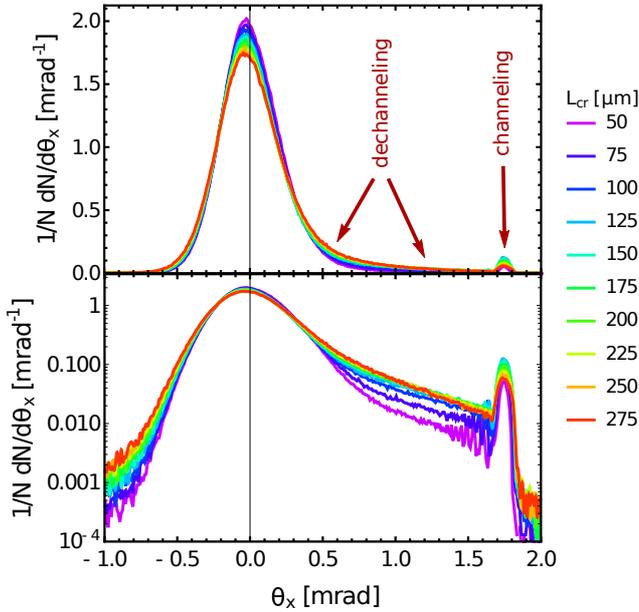}}
\caption{\label{Fig6} Angular distribution of the deflected beam as function of the crystal thickness for a Si crystal bent along the (111) planes with a deflection angle of 1.75 mrad in linear (top) and logarithmic (bottom) scale. The incoming beam distribution is the same as in Fig. \ref{Fig1}}
\end{figure}

The dependence of the deflection efficiency (an angular kick higher than 
$\theta_b-\theta_L$ to completely contain the channeling particles) on the 
crystal thickness is shown in Fig. \ref{Fig5}. The maximal efficiency is reached 
in the range of 120-150 $\mu m$ thickness. The physical explanation of 
the channeling effect considered above can be illustrated by the angular distribution of the deflected beam shown in Fig. \ref{Fig6} both in linear and logarithmic scale. 
One can see that for shortest thicknesses the particle fraction in the 
dechanneling tale is reduced as well as the number of particles not entered 
inside the angular acceptance increased, while for the longest ones the picture 
is inverse. This confirms our physical interpretation described above.

The maximal deflection efficiency is slightly above 1 \%. Though this number is 
considerably lower than the efficiency measured previously in the experiments 
\cite{PRL2014,EPJC2017,JINST2018,PRLSLACold,PRABSLAC,PRLSLAC,NIMBSLAC,
ModPhysSLAC}, it doesn't represents itself as an efficiency of extraction. 
In fact, not only channeling particles may reach the septum magnet entrance, but 
also the dechanneled ones. Moreover, the particles that did not enter under the 
channeling conditions can pass the crystal several times. In other words, the 
extraction efficiency is a multi-turn efficiency but not a single-turn one. 
Therefore, in order to calculate and optimize the extraction efficiency one 
needs to perform complete multi-turn simulations including particle dynamics in an 
accelerator.

\subsection{Radiation energy losses}

\begin{figure}
\resizebox{83mm}{!}{\includegraphics{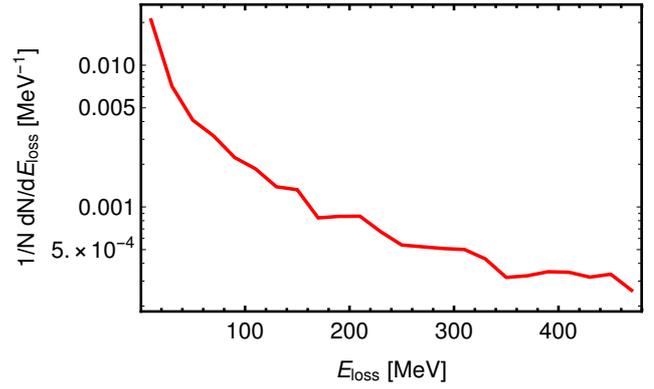}}
\caption{\label{Fig65} Radiation energy loss distribution for a 175 $\mu$m thick Si crystal, bent along the (111) planes with a deflection angle of 1.75 mrad. The incoming beam distribution is the same as in Fig. \ref{Fig1}}
\end{figure}

\begin{figure}
\resizebox{83mm}{!}{\includegraphics{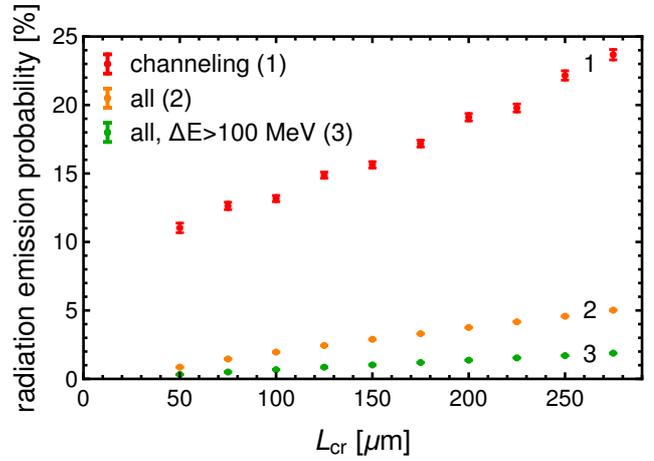}}
\caption{\label{Fig67} Radiation emission probability as function of the crystal thickness for (1) the channeled particles which were selected according to $\theta_x>\theta_b-\theta_L$ (red dots), for (2) all particles (yellow dots), and for (3) particles with radiation energy losses exceeding 100 MeV (green dots). The simulations were done for a Si crystal, bent along the (111) planes with a deflection angle of 1.75 mrad. The low-energy cut is 1 MeV.  The incoming beam distribution is the same as in Fig. \ref{Fig1}}
\end{figure}

\begin{figure*}
\resizebox{78mm}{!}{\includegraphics{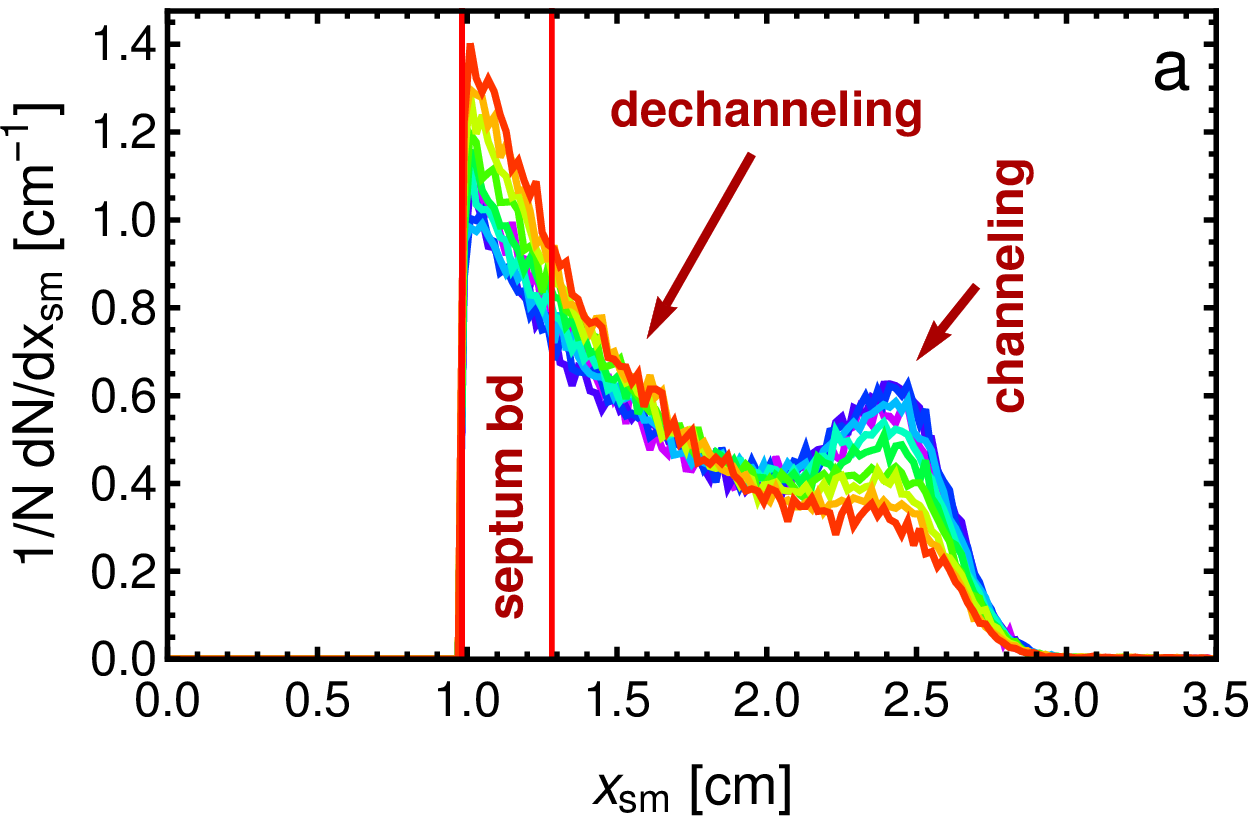}}
\resizebox{93mm}{!}{\includegraphics{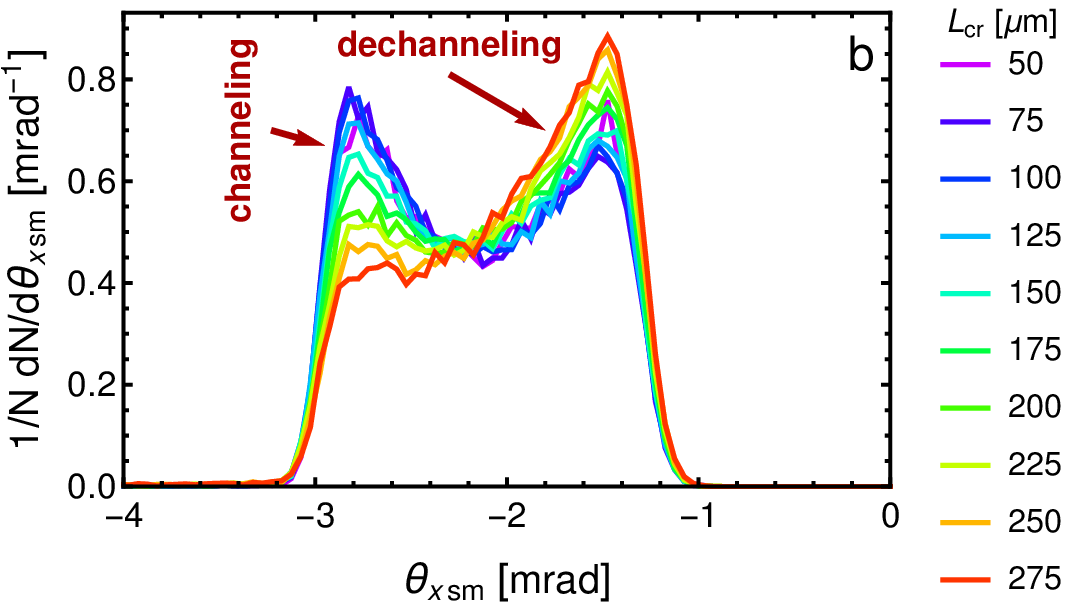}}

\caption{\label{Fig7} Transverse (a) coordinate and (b) angular distributions of the beam deflected by a bent crystal at the position of the septum magnet for different crystal thicknesses and Si crystals, bent along the (111) planes with a deflection angle of 1.75 mrad. The vertical lines in (a) indicate the septum magnet boundary. In this figure, only single crystal passage is taken into account}
\end{figure*}

\begin{figure}
\resizebox{83mm}{!}{\includegraphics{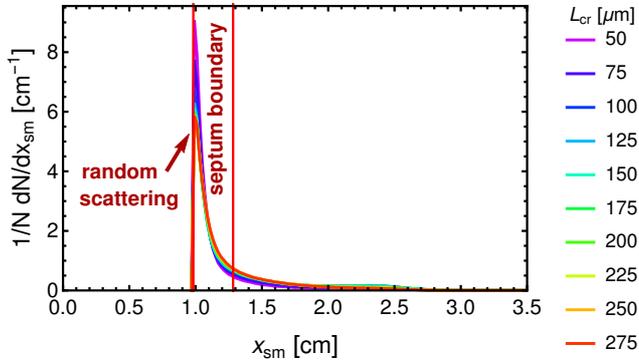}}
\caption{\label{Fig9} Multi-turn transverse coordinate distributions of the beam deflected by a bent crystal at the position of the septum magnet for different crystal thicknesses and Si crystals bent along the (111) planes with a deflection angle of 1.75 mrad after 100 revolutions in the accelerator. The vertical lines indicate the septum magnet boundary, both angular and energy cuts are applied}
\end{figure}

Radiation energy losses in a bent crystal can considerably modify particle 
trajectories in an accelerator. Hence it is very important to study 
both the radiation energy loss distribution and the radiation emission probability for the beam parameters and the crystal geometry considered in this paper.

An example of a radiation energy loss distribution in a 175 $\mu$m thick Si bent crystal with $\theta_b=$1.75 mrad is shown in Fig. \ref{Fig65} calculated for the incoming beam from Fig. \ref{Fig1}. The form of this distribution is similar to 
the bremsstrahlung spectrum. Though most of the particles lose few MeV, there is a 
non-negligible fraction exceeding 100 MeV, which will be directed outside the 
septum magnet (compare Fig. \ref{Fig4}). Therefore, it is important to estimate 
both the probability of radiation emission for the energies $>$100 MeV and the 
total probability, i.e. the entire range of energy losses. These results in 
dependence of the crystal thickness are shown in Fig. \ref{Fig67}.

The total radiation probability (orange in Fig. \ref{Fig67}) does not exceed 5 
\%, meaning a rather low fraction of particles though not negligible. The 
probability of radiation energy losses larger than 100 MeV (green in Fig. 
\ref{Fig67}) is less than 2 \%. Therefore, most of the particles are not affected 
by the radiation process and can pass the crystal again if they are not captured 
under channeling conditions. This makes multi-turn crystal-based 
extraction of electrons possible.

The total probability of radiation emission by channeling particles selected 
according to the definition of deflection efficiency (see Fig. \ref{Fig5}) is 
also shown in Fig. \ref{Fig67}. As can be seen, channeling radiation gives a considerable rise to the radiation emission probability, even exceeding 20 \% at 
higher crystal thicknesses. However, this does not modify strongly the 
deflection efficiency since the fraction of channeling particles that does not 
produce radiation is still high as shown in Fig. \ref{Fig6}. Moreover, the particles losing low enough energy can be also intercepted by the septum magnet.

Therefore, though the process of radiation emission in a bent crystal needs to be 
definitely taken into account for crystal-based electron beam extraction, its 
contribution should not considerably decrease the multi-turn extraction 
efficiency.

\subsection{Multi-turn simulations of crystal-based extraction}

The accelerator setup has been simulated with the CRYSTALRAD simulation code 
using the parameters indicated in Table \ref{tab:table1}-\ref{tab:table2}, the 
initial beam distribution at the bent crystal position from Fig. \ref{Fig1}, and the 
crystal parameters from Fig. \ref{Fig5}-\ref{Fig6} for different crystal 
thickness values. The crystal angular alignment was defined according to the 
center of the initial angular distribution in Fig. \ref{Fig1}b, being 0.97 mrad. The 
simulated coordinate and angular distribution at the septum magnet entrance after only one passage of the bent crystal are shown in Fig. \ref{Fig7}a-b, 
respectively. The channeling peak position on both 
distributions coincides with the peak position of an ideally deflected beam from Fig. \ref{Fig1}. However, a dechanneling fraction can be also intercepted by a septum magnet. All of these extracted particles need to be included in the extraction efficiency definition.

\begin{figure}
\resizebox{83mm}{!}{\includegraphics{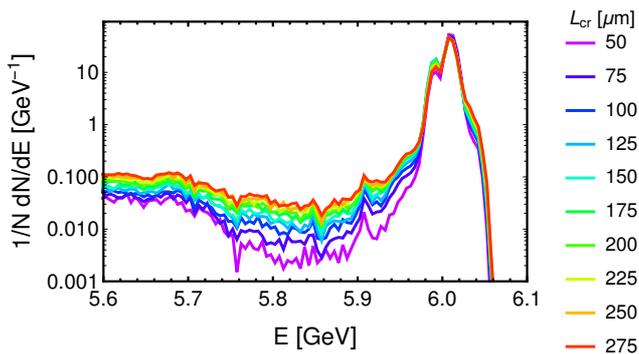}}
\caption{\label{Fig8} Energy distributions of the beam, deflected by a bent crystal at the position of the septum magnet for different crystal thicknesses and Si crystals bent along the (111) planes with a deflection angle of 1.75 mrad after 100 revolutions in the accelerator}
\end{figure}

\begin{figure*}
\resizebox{79mm}{!}{\includegraphics{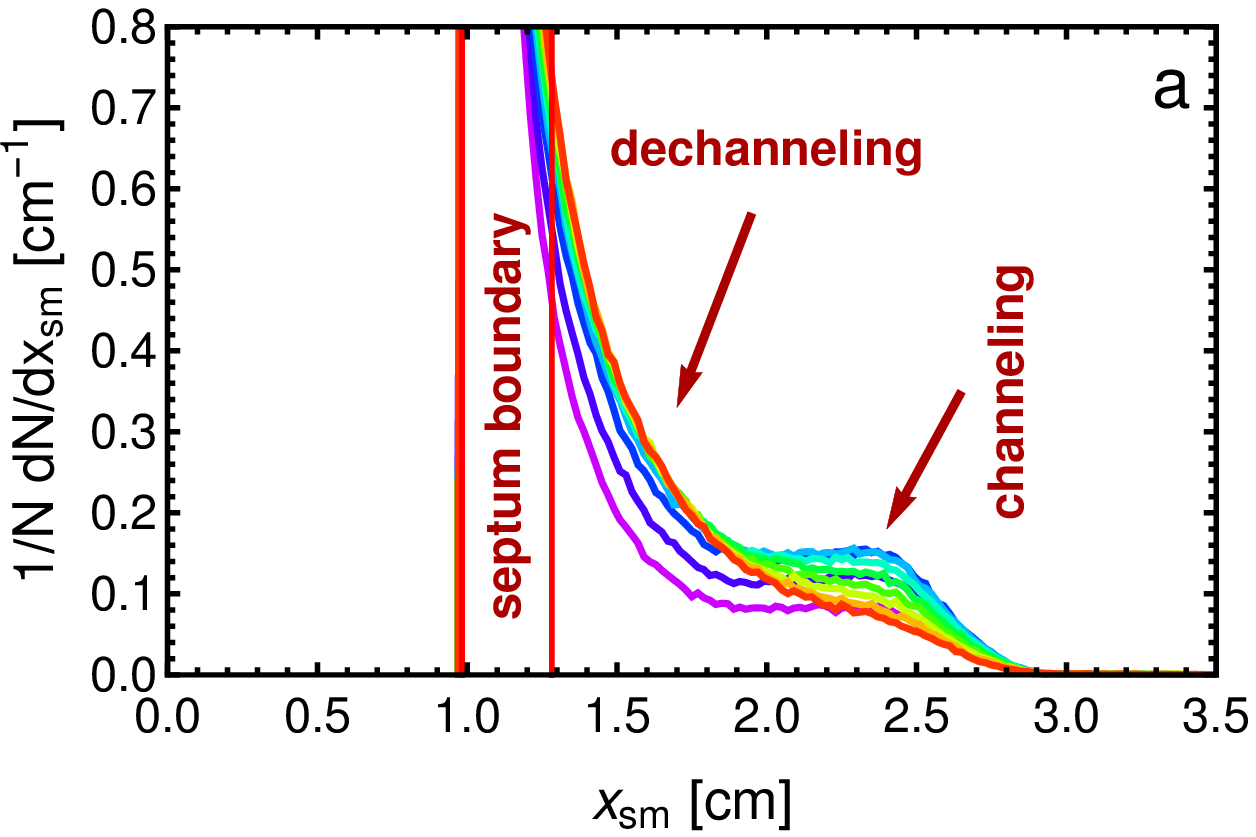}}
\resizebox{93mm}{!}{\includegraphics{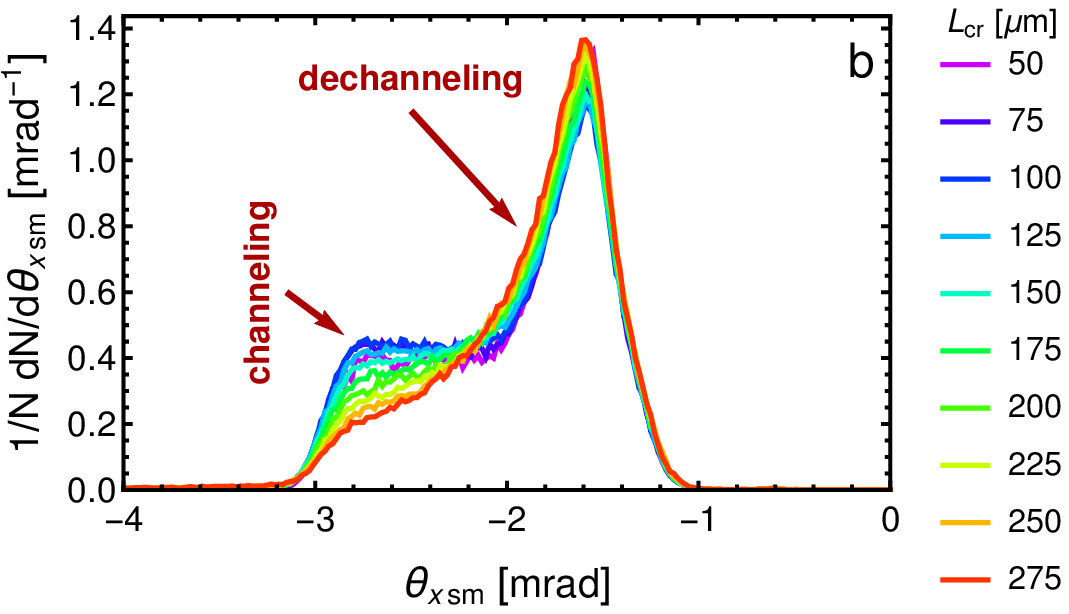}}

\caption{\label{Fig10}  Multi-turn (a) transverse coordinate and (b) angular distributions of the beam deflected by a bent crystal at the position of the septum magnet for different crystal thicknesses and Si crystals bent along the (111) planes with a deflection angle of 1.75 mrad after 100 revolutions in the accelerator. The vertical lines in (a) indicate the septum magnet boundary, all the cuts are applied}
\end{figure*}

\begin{figure}
\resizebox{83mm}{!}{\includegraphics{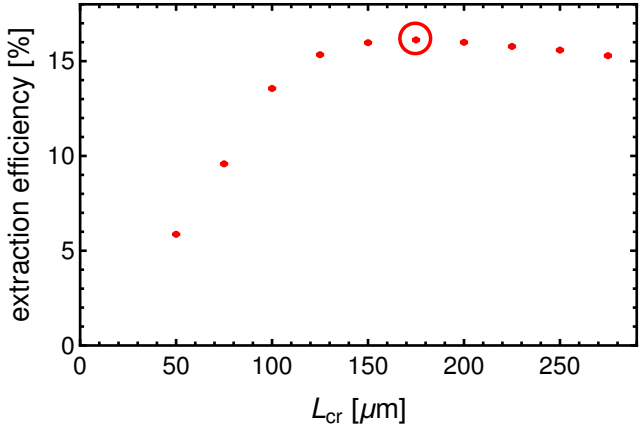}}
\caption{\label{Fig11} Extraction efficiency as function of the crystal thickness for Si crystals bent along the (111) planes with a deflection angle of 1.75 mrad. The red circle indicates the optimum case which is described in Table \ref{tab:table3}}
\end{figure}

A considerable part of the particles is randomly scattered by the crystal and is not 
captured under the channeling conditions as shown in Fig. \ref{Fig9}. 
The multiple scattering angle on the crystal lies in the range of $\sim 
50-90 \mu rad$, i.e. comparable with the critical channeling angle but is 
considerably lower than the deflection angle required. Consequently, most of randomly 
scattered particles do not reach the septum magnet entrance, but will be 
intercepted by its boundary. This fact confirms that the channeling effect is 
essential for the crystal-based extraction of an electron beam.

The extraction efficiency is defined as the ratio of the number of extracted particles over the number of particles at the crystal entrance. The number of 
extracted particles is calculated by using certain cuts applied to 
certain values at the end of the simulations, namely transverse coordinates, angles, 
energy, and number of turns in the accelerator. For practical reasons and based on the parameters of the extraction line, the cuts are chosen as described in the following paragraph.

The angular cut is defined to contain the beam in the range of [-4 ,0] mrad 
being in fact the whole extracted beam as shown in Fig. \ref{Fig7}b. The maximum number of turns in the accelerator is set to 100, since it is small enough to 
consider the beam energy as a constant, as well as high enough for charged 
particles to cross the crystal several times. The energy range follows from the 
particle energy distribution after 100 revolutions in the accelerator. It is 
presented in Fig. \ref{Fig8}. The energy cut is defined as 6 $\pm$ 0.1 GeV. It contains the main part of the beam within the RF-bucket as well as a fraction of particles losing less than 100 MeV. The coordinate cut is simply defined by the coordinates of the septum magnet boundary, as shown in 
Fig. \ref{Fig9} in the multi-turn coordinate distribution with all the remaining 
cuts applied. Namely, the coordinates must be $x_{sm} >$ 1.28 cm $\approx 5\,\sigma$.

Coordinate and angular distribution at the septum magnet entrance, taking into account the cuts as discussed beforehand, are shown in Fig. \ref{Fig10}a-b, respectively. Similarly to the single-passage case (compare Figs. \ref{Fig6}, \ref{Fig7}), 
there is an optimal value of the crystal thickness. In order to extract it, one 
needs to plot the extraction efficiency in dependence of the crystal thickness 
similarly to the deflection efficiency in Fig. \ref{Fig5}. This plot is 
shown in Fig. \ref{Fig11}. Qualitatively, there is a very similar dependence 
to that in Fig. \ref{Fig5} with the maximal efficiency point placed at 175 $\mu 
m$, i.e. being close to the optimal interval estimated above.

The main conclusion for the extraction efficiency in Fig. \ref{Fig11} concerns 
the maximum value of extraction efficiency that can be experimentally recorded 
reaching 16.1 \%. Namely, it is roughly one order of magnitude higher than the 
single-pass deflection efficiency. This is generally explained by the low multiple 
scattering angle w.r.t. the angular divergence and the low radiation emission 
probability as well. It means that if a particle is not captured under the channeling conditions during the first crystal passage, the initial conditions at 
the next ones may not be significantly different. Therefore, particles can be 
deflected into the septum magnet under the channeling conditions during the next 
crystal passages, however keeping in mind that the extraction septum is pulsed 
and its deflection is optimized for a certain beam energy. Nevertheless the 
multi-turn crystal-based extraction of electrons is possible and efficient enough.

\section{Bent crystal production}

\subsection{Technologies of bent crystal manufacturing}

Bent crystals are fabricated starting from Silicon-On-Insulator bonded wafers 
\cite{Wafers}. The high purity of these wafers is reached by means of melting-growing methods \cite{Growing}. The desired shape and thickness from few $\mu$m up to few mm can be obtained by depositing of a 100 nm layer of amorphous 
silicon nitride (Si3N4) onto both faces of the wafer through Low-Pressure 
Chemical Vapor Deposition followed by photolitography and anisotropic chemical 
etching of silicon \cite{Crystals2}.

\begin{figure}[!b]
\begin{center}
	\resizebox{44mm}{!}{\includegraphics{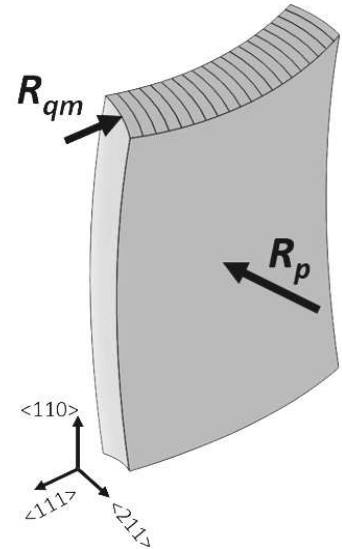}}
\end{center}
	\caption{\label{FigCrystalQM} Thanks to its mechanical anisotropic properties, bending of a silicon crystal along its main size to a radius $R_p$ leads to a secondary bending along its thickness, manifesting with a curvature of radius $R_{qm}$, which will be used to deflect charge particles by a bent crystal. <111>, <110> and <211> represent the directions of crystallographic axes}
\end{figure}

Once shaped, crystals are mounted on two plugs of a mechanical bender. Bringing both plugs closer together generates a bending moment that makes the crystal bent around 
the [111] direction. As shown in Fig. \ref{FigCrystalQM}, due to the so-called quasimosaic effect 
\cite{Quasimosaic1,Quasimosaic2}, this primary curvature $R_p$ produces a secondary 
bending of (111) planes $R_{qm}$ which will be used to deflect charge particles. A special technology of a dynamical holder has been developed \cite{EPJC2017,JINST2018} which allows to adjust the crystal curvature after the installation of the crystal into the accelerator vacuum system. This option offers the possibility to optimize the deflection angle during the experiment.

\subsection{Bent crystal characterization}

\begin{figure}[!b]
	\resizebox{83mm}{!}{\includegraphics{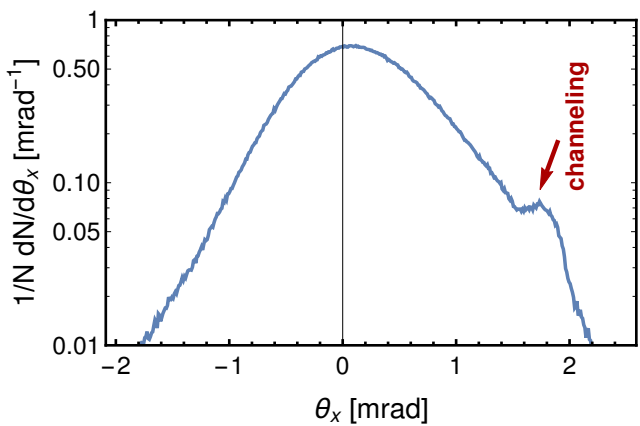}}
	\caption{\label{Fig14} Angular distribution of a deflected electron beam with an energy of 855 MeV, representing the beam conditions of the Mainz Mikrotron MAMI (University of Mainz, Germany), and a 175 $\mu$m thick Si crystal which is bent along the (111) planes with a deflection angle of 1.75 mrad}
\end{figure}

A preliminary characterization is done by using  high resolution X-ray diffraction, which will allow to reveal an inhomogeneity of the crystal bending and the crystal lattice mosaicity as well. Both factors in theory can significantly reduce the deflection efficiency. However, the modern technique of crystal manufacturing described before allows to produce bent Si crystals with almost zero mosaicity, without any significant imperfections and with very low inhomogeneity of the crystal curvature, though it hast to be verified for each crystal sample.

A unique way to characterize the crystal is to test it under real conditions, i.e. preferably at similar beam energies with ultra-relativistic electrons. For the planned experiment at DESY II, a suitable test could be performed by using the 855 MeV electron beam of the Mainz Mikrotron MAMI (University of Mainz, Germany). In principle the experimental setup described in Refs.~\cite{EPJC2017,EPJC2020,EPJC2021} could be re-used.
The low angular divergence of the MAMI beam of 21.4 $\mu$rad is advantageous because it is one order of magnitude smaller than the critical channeling angle for sub-GeV energies. For this test experiment and the crystal geometry optimized for DESY II, the simulated angular distribution of the deflected MAMI beam is shown in Fig.~\ref{Fig14}. As can be seen, the channeling peak which will serve as unique measure for the characterization of the channeling effect in the crystal can clearly be visible.

\section{Preliminary experimental setup design}

This section gives a brief overview of the experimental design considerations. 
The bent crystal will be mounted in an experimental chamber which is embedded in 
the extraction region into the former DORIS transfer line, c.f. Fig.~\ref{fig4}. 
The  optimized crystal parameters for the planned setup are summarized in 
Table~\ref{tab:table3}. The potential well correspoding to this parameters is shown in Fig. \ref{FigI1}b. In order to achieve the required precision for the 
crystal alignment w.r.t. the beam axis, it will be mounted onto a multi-axis 
remotely controlled goniometric stage.

\begin{table}
\centering
\caption{\label{tab:table3} Optimized crystal parameters for the planned experimental setup}
\label{parset}
\begin{tabular*}{\columnwidth}{@{\extracolsep{\fill}}ll@{}}
\hline
Parameter&Value\\
\hline
Bent crystal thickness& 175 $\mu m$\\
Bent crystal bending angle& 1.75 mrad\\
Bent crystal transverse position & -0.63 cm\\
Bent crystal angular alignment& 0.97 mrad\\
Septum magnet transverse position&  0.98 cm\\
\hline
\end{tabular*}
\end{table}

\begin{figure}[!t]		
	\centering
	\includegraphics[scale=0.31]{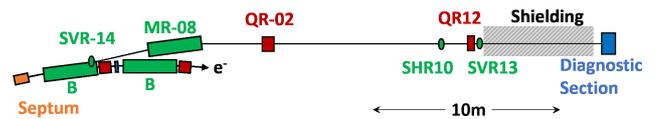}
	\caption{Overview of the former DORIS transfer line from the extraction region up to the diagnostic section which is located behind a shielding wall in the old transfer tunnel. The line consists of a bending  magnet (MR-08) and a pair of horizontally (QR-02) and vertically (QR12) focusing  quadrupoles. Fine alignment of the beam position is possible via steerer magnets in horizontal (SHR10) and vertical (SVR-14 and SVR13) direction. The direction of the circulating beam (e$^-$) is indicated by the arrow. The distance between the extraction point and the detector section is about 35 m}
	\label{fig19}
\end{figure}

\begin{figure}[!b]		
	\centering
	\includegraphics[scale=0.45]{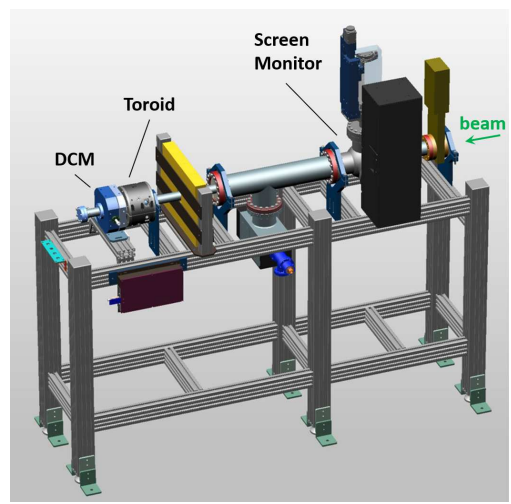}
	\caption{Diagnostic station for the detection of deflected and extracted beams from the DESY II accelerator. It consists of a screen monitor for the measurement of beam position and shape together with a beam transformer (toroid) and a dark current monitor (DCM) for bunch charge measurements. In this picture, the beam is coming from the right side}
	\label{fig20}
\end{figure}

The deflected beam will then be injected into the remaining part of the former 
transfer line which is depicted in Fig.~\ref{fig19}. It consists of a bending 
magnet (MR-08) and a pair of horizontally (QR-02) and vertically (QR12) focusing 
quadrupoles. Fine alignment of the beam position is possible via steerer magnets 
in horizontal (SHR10) and vertical (SVR-14 and SVR13) direction. About 35 m away 
from the extraction point there is a diagnostic station which is installed right 
behind a shielding wall separating the DESY II accelerator tunnel and the transfer channel. A drawing of this station which was recently commissioned is 
shown in Fig.~\ref{fig20}. In principle, a second diagnostic section could be 
refurbished which is right in front of the MR-08 dipole. However, at present it 
is assumed that the station behind the shielding wall will be sufficient for the 
commissioning of the experiment. 


This station includes a scintillating screen monitor for the measurement of the beam 
position and the shape which is described in Ref.~\cite{Wiebers:IBIC13-WEPF03}. If 
the beam hits the scintillator material, light is created in a multistage 
stochastic process with an intensity distribution which is proportional to the 
charge density of the incoming beam, resulting in a beam image that can be 
detected using a conventional area scan camera. In the present monitor setup, 
Lu$_{2(1-x)}$Y$_{2x}$SiO$_5$:Ce (LYSO:Ce) is used as scintillator material, and 
the intensity profile is detected by means of a standard CMOS camera. Using this 
type of scintillator it is possible to measure beam profiles for bunch charges 
down to the level of a few pC. For precise bunch charge measurements, a fast 
beam transformer (toroid) is used where the beam acts as primary single turn 
winding in a classical transformer such that it couples inductively to the 
measurement device. Details about the system which is able to detect bunch charges down to the level 1 -- 2 pC can be found in 
Refs.~\cite{Werner:DIPAC11-MOPD65,Werner:IBIC14-WEPF02}. For lower charges, the 
so-called dark current monitor (DCM) will be used. It consists of a resonator 
made from stainless steel with the frequency of the first monopole mode at 1.3 
GHz and a bandwidth of 6.7 MHz, for more details see 
Refs.~\cite{Lipka:DIPAC11-WEOC03,Lipka:IBIC13-WEPF25}. Using this kind of 
device, in principle it is possible to measure bunch charges down to the level 
of 20 fC.

Using the available beam monitor devices it is possible to optimize the crystal 
parameters and transfer line settings based on the measured intensity of the 
deflected beam. In principle it can also be considered to measure the phase 
space of the extracted beam. To do so, the screen 
monitor can be used in combination with the two quadrupoles in order to 
perform a quadrupole scan based emittance measurement, see e.g. 
Ref.~\cite{Minty}. However, it still has to be investigated if the detection 
sensitivity, the phase advance of both quadrupoles and the stability of the extracted 
beam will allow such type of measurements.

\section{Conclusions}

The first design of a crystal-based extraction of 6 GeV electrons from the DESY 
II booster synchrotron has been proposed. The main idea of this technique is to 
extract beam halo from an accelerator in a parasitic mode based on the 
application of the channeling effect in a bent crystal. 

The experimental setup will include a bent crystal which is located in the 
extraction section of the synchrotron, just behind the extraction kicker and 
followed by the septum magnet and a transfer line equipped with beam diagnostic monitors as described in the previous section. A bent crystal placed at a horizontal offset of 3$\sigma$ will 
deflect beam halo onto the septum magnet. This deflection can be supported by 
the beam kicker which will shift the beam towards the septum magnet. Both 
crystal-to-beam and septum-to-beam positions can be adjusted. 

The beam distribution at the crystal entrance has been simulated. A crystal bending 
angle of 1.75 mrad has been chosen to ensure the extraction of charged 
particles under the channeling conditions.

A multi-turn extraction process has been studied using the CRYSTALRAD 
simulation code, taking into account radiation processes. It has been shown 
that though the radiative energy loss may kick out charged particles from an 
extraction trajectory, the single-turn extraction efficiency will not be affected much 
and may reach 1 \% as well. Moreover, since the average radiation probability 
does not exceed 5 \%, multi-turn extraction efficiency is not considerably affected by the radiation process.

The extracted beam distribution has been simulated. The efficiency 
of beam extraction within $\pm 0.1$ GeV and $\pm 2$ mrad of energy and 
angular distribution respectively has exceeded 16 \%.

In general, due to the availability of the beam extraction only for a short 
moment and at a certain energy while the booster synchrotron is ramping up its 
energy, the channeling setup at DESY II is considered more as a 
proof-of-principle experiment for future applications than as an alternative 
for the existing internal targets. Nevertheless, the future experimental tests 
of the electron crystal-based extraction technique might raise interest because 
it can be applied at existing electron synchrotrons worldwide in a 
straightforward manner. This will make high quality intense electron beams more 
accessible for a larger community, and consequently, will considerably speed up 
nuclear and particle physics detectors and generic detector R\&D, as well as 
will be very useful for many projects in high-energy physics requiring 
fixed-target experiments. Some applications of crystals as X-ray and gamma sources \cite{PRL2015,EPJC2021,PRA2012,NIMB2013,PRL2013,ICHEP2016,PRL2018,BandieraNIMA,NIMBSLAC,ModPhysSLAC,CU1,CU2,CU3,CU4}, compact electromagnetic calorimeters \cite{PRL2018,BandieraNIMA,KLEVER} and positron sources \cite{Positron1,Positron2,Positron3,Positron4} could be also tested in these experiments. Furthermore, since FCC-ee is becoming one of the main 
after-LHC collider projects, electron/positron crystal-based extraction may 
provide an access to unique experimental conditions for ultra-high energy 
fixed-target experiments to measure e.g. CP-violation processes and physics 
beyond the Standard Model.

\begin{acknowledgements}
We acknowledge partial support of the INFN through the MC-INFN and the STORM projects. A. Sytov acknowledges support by the European Commission (the TRILLION project within the H2020-MSCA-IF-2020 call, GA. 101032975). A. Romagnoni acknowledges support from the ERC Consolidator Grant SELDOM G.A. 771642. We also acknowledge the CINECA award under the ISCRA initiative for the availability of high performance computing resources and support.
\end{acknowledgements}

\end{document}